\documentstyle[12pt,cite,epsfig,psfrag]{article}

\def\be{\begin{equation}}
\def\ee{\end{equation}}
\def\bea{\begin{eqnarray}}
\def\eea{\end{eqnarray}}

\newcommand{\beqal}{\begin{eqnarray}\label}
\newcommand{\beqa}{\begin{eqnarray}}
\newcommand{\eeqa}{\end{eqnarray}}

\begin{document}
\baselineskip=.6cm
\begin{titlepage}
\begin{center}
\vskip .2in

{\Large \bf  Some New Light-cone Time Dependent Solutions in Deformed 
pp-wave Backgrounds}
\vskip .5in

{\bf Srikumar Sen Gupta \footnote{e-mail: srikumar@iopb.res.in}
}\\
\vskip .1in
{\em Institute of Physics,\\
Bhubaneswar 751 005, INDIA}

\end{center}

\begin{center} {\bf ABSTRACT}
\end{center}
\begin{quotation}\noindent
\baselineskip 15pt
Brane solutions in time\,/\,light-cone time dependent backgrounds are of 
interest in order to gain a deeper understanding of the physics associated 
with cosmological and null singularities. 
In this paper, we report 
both brane solutions and their bound states in light-cone time dependent pp 
wave-like backgrounds. We show that the backgrounds our solutions live in are 
all cosmologically singular (lightlike) at the classical level. A spacetime 
supersymmetry analysis of the backgrounds reveals that they retain 1/8 of the 
full type IIB supersymmetry. 
\end{quotation}
\vskip 3.6in
September, 2013\\
\end{titlepage}
\vfill
\eject

\section{Introduction}
The study of time-dependent backgrounds in the context of string theory is of great importance because of its obvious relevance to cosmology, and, more importantly, because of the hope it holds out in the resolution of black-hole and cosmological singularities, such as the big-bang singularity, which cannot be resolved otherwise within classical general relativity. And, to focus on a subset of some recent attempts, one may cite, for example, the time-dependent orbifold models \cite{LMS1,LMS2,CC1,CC2}. These models are solvable perturbatively and they preserve some of the supersymmetries. However, owing to blue shifting of modes in these backgrounds, these models are unstable to large back reaction \cite{HP,L,BCKR}. Sen's idea of the rolling of the open string tachyon on an unstable brane (brane anti-brane pair) provides another example of a solvable time-dependent model \cite{AS}. And, in \cite{MS} the authors have argued in favour of replacing 
the cosmological singularity by a closed string tachyon condensate. By so doing it is possible to truncate self-consistently the perturbative string amplitudes to the small coupling region such that they are finite, suggesting, thereby, a resolution of the singularity. 
\\

Yet another exciting proposal to hit this field in recent years is the idea of matrix big bang \cite{CSV}. In this remarkable paper the authors work with type II-A theory in a linear null dilaton background which retains one-half of the original supersymmetries. Near the big bang or the big crunch the linear null dilaton makes the string coupling very large and this leads to the breakdown of perturbative string theory. As a way out, a dual matrix string theory has been proposed and the physics near the big bang/crunch  is described by a weakly coupled two dimensional super Yang-Mills theory on the Milne orbifold. Hence, the matrix degrees of freedom, rather than the point particle or the perturbative string, correctly describe the physics in the vicinity of the singularity. The following works, often drawing upon various holographc dual ideas, furnish further generalisations of the above proposition ( \cite{DMNT} and references therein for a partial list of such studies).   \\

Thus, in the backdrop of the above interesting developments, it is important 
that one searches for new time-dependent backgrounds in string theory and 
supergravity. Of particular interest are the solutions whch enjoy partial 
supersymmetry (BPS solutions) because they allow us a peek into those regions 
where non-perturbative physics holds sway. Supergravity backgrounds merit 
attention too because they help us speculate on the corresponding dual field 
theory and they often offer us valuable insight into time-dependent sources in 
gauge theory (see, for example, \cite{DMNT}, among other such studies). However, 
obtaining explicit solutions of such types in 
general has proven to be difficult. In this paper we continue our search 
for new solutions in geometries akin to the usual pp-wave backgrounds. We 
do this by  constructing a D-string solution in a light-cone time dependent 
pp-wave-like background. The solution is then used to generate both Dp brane 
and bound state solutions  in analogously deformed pp wave backgrounds.   \\

The layout of the paper is as folows. In section (2) we present our 
background which is a deformation of the RR $AdS_3\times\,S^3\times R^4$ pp wave 
background, the deformation being a function of light-cone time. 
We propose our ansatz for a D-string solution in this background. 
The solution can then be used to seed other Dp-brane solutions as well as 
D-brane bound state solutions in both NS-NS and RR (deformed) pp wave 
backgrounds. 
Of the 
many possibilities we choose to construct a D3 brane solution and a D1-D3 bound 
state solution and we draw attention to the similarity of structure that exists 
between our solutions and the ones found in the literature and we point out a 
significant difference as well. In section (3) we focus our attention on the 
backgrounds in which our solutions live and we use various classical GR tests to 
establish the fact that all our backgrounds are cosmologically singular at the 
classical level. Section (4) discusses the supersymmetry of our D-string 
solution and shows that the solution preserves 1/8 of the full type IIB supersymmetry. Since the other two solutions are derived from the parent D-string solution by application of T-duality and since supersymmetry is preserved by T-duality, they too preserve the same amount of supersymmetry. Finally, we summarize our results in section (5).


\section{Dp brane and bound state solutions in light-cone time dependent pp-wave backgrounds}
\noindent

In this section we first present our ansatz for a supergravity background which 
depends on $x^+$, a light-like coordinate .We choose to work in 10-dimensional 
string frame in type IIB supergravity. Our bakground is supported by a five form 
field  and we assume that only the metric and the dilaton are excited :
\begin{eqnarray}
ds^2&=&f^{-{1\over2}}_{1b}(2 dx^+dx^- - \mu^2{\sum_{i=1}^{4}}{x^i}^2(dx^+)^2)
+ f^{1\over2}_{1b}{\sum_{a=1}^{8}}(dx^a)^2  \cr
& \cr
e^{2\phi}&=& f_{1b},~~~~F_{+1256} = F_{+1278} = F_{+3456} = F_{+3478} = 2\mu, 
\label{D-string background}
\end{eqnarray}
where $ f_{1b}$ depends on the light-like coordinate $x^+$ i.e. $f_{1b}=f_{1b}(x^+)$.
Now, viewing $x^+$ as light-cone time, our background can be interpreted as a light-cone-time dependent deformation of the pp-wave background originating from the Penrose limit of R-R $AdS_3\times\,S^3\times R^4$. The supergravity solution for a D-string in such a background is then given by :
\begin{eqnarray}
ds^2&=&f^{-{1\over2}}_1(2 dx^+dx^- - \mu^2{\sum_{i=1}^{4}}{x^i}^2(dx^+)^2)
+ f^{1\over2}_1{\sum_{a=1}^{8}}(dx^a)^2  \cr
& \cr
e^{2\phi}&=& f_1,~~~~F_{+1256} = F_{+1278} = F_{+3456} = F_{+3478} = 2\mu, \cr
& \cr
F_{+-a}&=&\partial_a f^{-1}_1,~~~~f_1 = f(x^+) + {Q_1\over r^6}
\label{D-string}
\end{eqnarray}
where $f_1$ satisfies the Green's function equation in the eight dimensional 
transverse space of the D-string and $Q_1$ is a constant.
We have verified that the above solution satisfies the type-II B field equations (see, e.g., \cite{DKL},\cite{AIO}). In particular, we point out that the $\mu$ dependence appears only in the equation involving the $R_{+\,+}$ component of the Ricci tensor and it is cancelled by contributions from the dilaton and the five-form field strength. It also folows from the same equation that the background is a solution provided the following condition holds :

\begin{equation}
{ {f_1,}_{+\,+}\over{f_1}} =  0.
\label{condition}
\end{equation}
which implies that $f_1= A\,{x^+}+B$, where the integration constants A and B 
are independent of $x^+$. $f_1$ is thus a linear function of $x^+$. Comparing 
this with our ansatz for $f_1$ we conclude that $f(x^+) = A\,x^+$ and $B= {Q_1\over r^6}$. And, since in the asymptotic limit of equation (2) one 
recovers the background given by (1), it follows therefore that 
$f_{1b}=f(x^+)=A\,x^+$. A consequence of this linear dependence of $f_{1b}$ 
on $x^+$ is that as  $x^+\,\rightarrow 0$ the transverse 
dimensions $x^i$ (i = 1,2,.....,8) shrink to zero size. This, then, implies an 
approach to the big bang point in so far as the transverse coordinates are 
concerned. Indeed, we have shown in section 3 that the background has a 
cosmological (lightlike) singularity at $x^+=0$. Furthermore, the dilaton 
associated with the background is a logarithmic function of $x^+$ 
(i.e.\,$\phi=\frac{1}{2}$\,$\log\,(A\,x^+)$), defined only for $x^+\geq0$. The dilaton, therefore, blows up as $x^+\,\rightarrow 
0$  and the string coupling given by $g_s=e^\phi$ goes to zero and vanishes at the singularity. String coupling is thus weak close to the big bang and strong at late times and, with all the transverse dimensions zero-sized, it is not immediately clear whether perturbative string theory can still be relied on\cite{LI}. Seen from this point of view, we observe that our background bears resemblance to the ones studied in\cite{DMNT,LI} and stand in striking contrast to some other backgrounds found in the literature, where the string coupling diverges as the singularity is approached.    \\
Now, solution (\ref{D-string}) can be used to generate other Dp-brane solutions $\,(p = 1,3,..,5)$ in both R-R and NS-NS light-cone time dependent pp-wave backgrounds by apropriate applications of T and S duality transformations, following the ten-dimensional T-duality map between the type IIA and the type IIB string theories as given in \cite{BMM,BHO} and \cite{S} (for S duality).  For example, applying two successive T dulities along $x^5$ and $x^6$ we obtain a D3 brane solution of the form : 
\begin{eqnarray}
ds^2&=&f^{-{1\over2}}_3(2 dx^+dx^- -\mu^2{\sum_{i=1}^{4}}{x^i}^2(dx^+)^2+
(dx^5)^2+(dx^6)^2)\cr
& \cr
&+& f^{1\over2}_3{\sum_{a=1,..,4,7,8}}(dx^a)^2  \cr
& \cr
F_{+12}& =&F_{+34} = - 2\mu, \cr
& \cr
F_{+ - 5 6 a}&=& \partial_a {f_3}^{-1},~~~~e^{2\phi} = 1,
\label{D3-brane}
\end{eqnarray}
where $f_3$ is the harmonic function in the six dimensional transverse space of 
the D-3 brane. The procedure, of course, involves de-localization of the brane 
along the directions in which T-duality is implemented. One can check that the 
above solution satisfies the type IIB field equations provided $f_3$ satisfies a 
condition similar to (\ref{condition}) making $f_3$ too a linear function 
of $x^+$. T-duality can be further exploited to produce configurations 
involving D($p+1,p-1$)-brane bound states(see \cite{BMM1,RRN,NOKLP,NOKLPSS} for 
a representative sample of more such studies). 
To provide an example of how a bound state solution can be constructed out of solution (\ref{D-string}), we first apply a T duality along $x^5$ to produce a D2 brane stretching along the $x^+$,\,$x^-$,\,and $x^5$ directions. The D2 brane is also delocalized along one of the transverse directions, say $x^6$. The $x^5$-$x^6$ plane is then rotated through an angle $\varphi$ about the $x^-$ axis to bring them to new orientations which now define a new pair of orthogonal axes $\tilde x$-$\tilde y$. In this new configuration the original D-2 brane now lies tilted at an angle $\varphi$.  Finally, the D2 brane is T-dualized along $\tilde x$ to yield a D1-D3 bound system of the form :

\begin{eqnarray}
ds^2&=&X^{-{1\over2}}(2 dx^+dx^- - \mu^2{\sum_{i=1}^{4}}{x^i}^2(dx^+)^2)
+ X^{1\over2}[\,{ d \tilde x^2 + d \tilde y^2 \over 1 + (X-1) \cos^2 \varphi}+\!                            {\sum_{a=1,..,4,7,8}}(dx^a)^2]
\nonumber\\
e^{2\,\phi} &=&\, X \over  1 + (X-1) \cos^2 \varphi
\nonumber\\
H_{\tilde x \,\tilde y \,a} &=&\partial_a\, [\, { (X -1) \cos \varphi\, \sin \varphi \over  1 + (X-1)  
\cos^2 \varphi}\,]
\nonumber\\
\chi&=&0
\nonumber\\
F^{(3)}_{+\,-\,a}&=&\partial_a\, [\,{ \sin \varphi \over X}\,]
\nonumber\\
\nonumber\\
F^{(3)}_{+\,1\,2}&=&-2\mu\,cos\varphi
\nonumber\\
\nonumber\\
F^{(3)}_{+\,3\,4}&=&-2\mu\,cos\varphi
\nonumber\\
F^{(5)}_{+ - \,\tilde x \,\tilde y\, a}&=&\partial_a\,[\,-{\cos \varphi \over X} +{1\over 2}\, ({{1-X}\over X}){ \sin^2 \varphi \cos \varphi\, \over  1 + (X-1) \cos^2 \varphi}\,] 
\nonumber\\
\nonumber\\
F^{(5)}_{+ 1\,2 \,\tilde x \,\tilde y}&=&-\,\mu\,\sin \varphi\,[\, {{2+(X -1) \cos^2} \varphi\over  1 + (X-1)  \cos^2 \varphi}]
\nonumber\\
\nonumber\\
F^{(5)}_{+ 3\,4 \,\tilde x \,\tilde y}&=&-\,\mu\,\sin \varphi\,[\, {{2+ (X -1) \cos^2 \varphi}\over  1 + (X-1)  \cos^2 \varphi}]
\nonumber\\
\label{D1-D3}\
\end{eqnarray}
where X satisfies the Green's function equation in the six dimensional 
transverse space defined by $x^1$,....$x^4$, $x^7$, and $x^8$. It can be checked 
that the above solution solves the type IIB field equations with X being a 
linear function of $x^+$. We observe that the solution involves both $F^{(3)}$ 
(i.e.$F^{(3)}_{+-a}$) and $F^{(5)}$ which require the presence of both a 
D-string stretching along ($x^+$, $x^-$) and a D-3 brane, the world volume of 
which includes the $x^+$, $\tilde x$, $\tilde y$, and the $x^-$ directions, to 
couple to them. The solution, therefore, represents a D1-D3 bound system. It is 
easy to see that with $\varphi=0$, one recovers the D-3 brane solution with 
$F^{(3)}_{+-a}$, $F^{(5)}_{+ 1\,2 \,\tilde x \,\tilde y}$, and$F^{(5)}_{+ 3\,4 
\,\tilde x \,\tilde y}$ all zero. This is because the D-2 brane has now been 
essentially T-dualized along the $x^6$ direction. We conclude this exercise by 
remarking that it is possible to generate more exotic configurations by 
involving T-duality along both transverse and world volume directions of some 
parent brane structure.\\

Finally, we wish to mention by way of an observation that the solutions so 
obtained are exactly similar in structure to the static solutions existing in 
the literature save for the fact that the static solutions have a constant part 
in the Green's functon in place of $f(x^+)$ as in our D-string ansatz ( and 
also in the other solutions obtained therefrom). In this context, it is also 
worth mentioning that in \cite{SSG} we found out, by an entirely different 
procedure, brane solutions in time-dependent backgrounds in M-theory as well as 
in type-II string theories and the solutions, quite interestingly, share the 
same feature as that mentioned above - namely, all these solutions have the constant 
part of the Green;s function replaced by a time-dependent one. This parallelity of 
structure between solutions originating from quite different contexts embolden 
us to suggest, without proof, a general prescription for obtaining brane 
solutions in time\,/\,light-cone time dependent backgrounds - all that 
one has to do is simply replace the constant part of the Green's function of a static Dp- brane/bound state solution by a suitable time-dependent function and, in so doing, one ends up with a Dp-brane/bound state solution in a time-dependent background.

\section{Study of the backgrounds supporting brane and bound system solutions }
\noindent
We now proceed to study the geometric features of the backgrounds in which our brane and bound state solutions live. As mentioned earlier, the solutions and backgrounds presented so far are all in the string frame. Now, the Einstein frame differs from the string frame by a function of the dilaton field 
\begin{equation}
ds^2_{E}= e^{-\phi /2} ds^2_{string}\,,
\end{equation}
giving rise to the possibility that a non-trivial dilaton may cause a singularity in the Einstein frame to disappear in the string frane \cite{CH}. The subtler nature of the Einstein frame, therefore, qualifies it to capture the geometry of a spacetime better\cite{CHZ}. We therefore study our backgrounds in the Einstein frame. The backgrounds are:

1. D-string background :
\begin{eqnarray}
ds_E^2&=&f^{-{3\over4}}_{1b}(2 dx^+dx^- - \mu^2{\sum_{i=1}^{4}}{x^i}^2(dx^+)^2)
+ f^{1\over4}_{1b}{\sum_{a=1}^{8}}(dx^a)^2  \cr
& \cr
e^{2\phi}&=& f_{1b},~~~~F_{+1256} = F_{+1278} = F_{+3456} = F_{+3478} = 2\mu, \cr
& \cr
f_{1b} &=& f(x^+)=A\,x^+
\label{E-frame D-string background}
\end{eqnarray}

2. D-3 brane background :
\begin{eqnarray}
ds_E^2&=&f^{-{1\over2}}_{3b}(2 dx^+dx^- -\mu^2{\sum_{i=1}^{4}}{x^i}^2(dx^+)^2+
(dx^5)^2+(dx^6)^2)\cr
& \cr
&+& f^{1\over2}_{3b}{\sum_{a=1,...,4,7,8}}(dx^a)^2  \cr
& \cr
F_{+12}& =&F_{+34} = - 2\mu, \cr
& \cr
e^{2\phi} &=& 1
\label{d3-brane background}
\end{eqnarray}

3. D1-D3 background :
\begin{eqnarray}
ds^2_E&=& [{1 + (X_b-1) \cos^2 \varphi \over {X_b}^3}]^{1\over4}\,(2 dx^+dx^- - \mu^2{\sum_{i=1}^{4}}{x^i}^2(dx^+)^2)\cr
&\cr
&+& [(1+ (X_b-1) \cos^2 \varphi)X_b]^{1\over4} [\,{ d \tilde x^2 + d \tilde y^2 \over 1 + (X_b-1) \cos^2 \varphi}+\!             {\sum_{a=1,..,4,7,8}}(dx^a)^2]
\nonumber\\
e^{2\,\phi} &=&\, X_b \over  1 + (X_b-1) \cos^2 \varphi
\nonumber\\
H_{\tilde x \,\tilde y \,a} &=&\partial_a\, [\,{ (X_b -1) \cos \varphi\, \sin \varphi \over  1 + (X_b-1)  
\cos^2 \varphi}\,]
\nonumber\\
\chi&=&0
\nonumber\\
F^{(3)}_{+\,-\,a}&=&\partial_a\, [\,{ \sin \varphi \over X_b}\,]
\nonumber\\
\nonumber\\
F^{(3)}_{+\,1\,2}&=&-2\mu\,cos\varphi
\nonumber\\
\nonumber\\
F^{(3)}_{+\,3\,4}&=&-2\mu\,cos\varphi
\nonumber\\
F^{(5)}_{+ - \,\tilde x \,\tilde y\, a}&=&\partial_a\,[\,-{\cos \varphi \over X_b} +{1\over 2}\, ({{1-X_b}\over X_b}){ \sin^2 \varphi \cos \varphi\, \over  1 + (X_b-1) \cos^2 \varphi}\,] 
\nonumber\\
\nonumber\\
F^{(5)}_{+ 1\,2 \,\tilde x \,\tilde y}&=&-\,\mu\,\sin \varphi\,[\, {{2+(X_b -1) \cos^2} \varphi\over  1 + (X_b-1)  \cos^2 \varphi}]
\nonumber\\
\nonumber\\
F^{(5)}_{+ 3\,4 \,\tilde x \,\tilde y}&=&-\,\mu\,\sin \varphi\,[\, {{2+ (X_b -1) \cos^2 \varphi}\over  1 + (X_b-1)  \cos^2 \varphi}]
\nonumber\\
\label{D1-D3 background}
\end{eqnarray}
Note that in all of the above the suffix 'b' denotes - ``background'', and both $f_{3b}$ and $X_b$ are, like $f_{1b}$, linear functions of time. We now employ some classical GR tests to look for possible singular behaviour of our background spacetimes. Since the D1-D3 system embodies the richest structure we analyze this background in detail. This background is found to be Ricci-flat ($R=0$), and $R^2\,(\equiv{R^{AB}R_{AB}})$ is zero too.. The only non-vanishing $R_{AB}$ of this background is $R_{++}$ (\,given in the appendix\,) and the gauge invariant curvature defined along a geodesic for which $x^-$, $x^a$, $\tilde x$, and $\tilde y$ are all constants is 
\begin{eqnarray}
R_{\lambda\lambda}(\lambda) \equiv R_{\mu\nu}\,{dx^{\mu}\over d{\lambda}}\,{dx^{\nu}\over d{\lambda}} &=& R_{++}\,({d\,x^+\over d{\lambda}})^2  \quad \quad \nonumber\\
&=&{\rm const}\,R_{++}\, \vert \frac {Q({x^+})^2+P\,x^+}{(1+\frac {P}{Qx^+})^2} \vert ^{\frac{1}{2}}                   \nonumber\\
\label{gauge invariant curvature}
\end{eqnarray}
\noindent
(see below for definitions of $P$ and $Q$ and also for a calculation of the quantity $({d\,x^+\over d{\lambda}})^2$). 

Now,  with $X_b = A x^+$ as in the case of our D-string solution, it is easy to 
see that as $x^+ \rightarrow 0$, $R_{++}$ goes as $(x^+)^{-2}$ and the 
expression within the modulus sign goes as $(x^+)^{\frac {3}{2}}$. Hence, $ 
R_{\lambda\lambda}(\lambda) $ grows as $\frac {1}{\sqrt x^+}$. Therefore, the 
gauge invariant curvature blows up at $x^+=0$. This hints at the existence of a 
lightlike curvature singularity at $x^+=0$. Since in GR a spacetime is 
singular if it is geodesically incomplete and cannot be embedded in a larger 
spacetime \cite{C}, it is worth delving into this aspect of our background.

The equation for a time-like geodesic in this background at constant $x^-$, $x^a$, $\tilde x$, and $\tilde y$, i.e. one moving only along $x^+$, is given by
\begin{equation}
\frac{d^2 x^{+}}{d\lambda^2} +
\Gamma^{+}_{++}(\frac{dx^+}{d\lambda})^2 = 0,
\label{geodesic eqn}
\end{equation}
where
\begin{equation}
\Gamma^{+}_{++} = -\frac{1}{4}\frac{2\,Q\,x^++3P}{Q({x^+})^2+P\,x^+},
\end{equation}
and\,${\rm P}={\rm sin}^2\varphi$, ${\rm Q}={\rm A}\,{\rm cos}^2\varphi$. Now, plugging this into equation (\ref{geodesic eqn}), we get
\begin{equation}
\frac{d^2 x^+}{d\lambda^2} -[\frac{1}{4}\frac{2\,Q\,x^++3P}{Q({x^+})^2+P\,x^+}]
(\frac{dx^+}{d\lambda})^2 = 0. 
\label{geodesic eqn-1}
\end{equation}


The following auxiliary functions, $y(\lambda)$ and $p^{\prime}(x^+)$, are now introduced with a view to solving (\ref{geodesic eqn-1}) :
\begin{eqnarray}
y(\lambda) &\equiv& {dx^+\over d\lambda},\nonumber\\
p^{\prime}(x^+) &\equiv&  {dp\over dx^+}  = -\frac{1}{4}\frac{2\,Q\,x^++3P}{Q({x^+})^2+P\,x^+}.
\label{auxiliary functions}
\end{eqnarray}
\noindent

The above functions help us to reduce eq.\,(\ref{geodesic eqn-1}) to
\begin{equation}
{dy\over d\lambda}\,+\,p^{\prime}(x^+)\,y^2 = 0.
\label{modified geodesic eqn}
\end{equation}
\noindent
We then solve (\ref{modified geodesic eqn}) to obtain the following expression for $\lambda$ in terms of $p(x^+)$ :
\begin{equation}
\lambda = {\rm const}\,\int e\,^{p(x^+)}\,dx^+.
\label{lambda soln }
\end{equation}
\noindent
The solution entails obtaining the following intermediate step which we make use of in the calculation of gauge invariant curvature above :
\begin{equation}
y(\lambda) \equiv {dx^+\over d\lambda} = {\rm const}\:e^{-p}.
\label{intermediate eqn}
\end{equation}
\noindent
Again, it follows from (\ref{auxiliary functions}) that the solution for $p(x^+)$ is
\begin{eqnarray}
p(x^+) &=&  \ln \vert \frac {Q({x^+})^2+P\,x^+}{(1+\frac {P}{Qx^+})^2} \vert ^{-\frac{1}{4}} + {\rm const}.
\label{p soln}
\end{eqnarray}
\noindent




It follows now from (\ref{lambda soln }) that

\begin{equation}
\lambda = {\rm const}\, (\frac{P}{Q^2})^{1\over 4} \int \vert \frac{1}{(x^+)^3}\,(1+\frac{Q}{P}x^+)\vert
^{\frac{1}{4}}\,dx^+.
\label{lambda soln 1}
\end{equation}
\noindent

Clearly, this integral is undefined at $x^+=0$. So, to investigate its behaviour at $x^+=0$, we evaluate the above indefinite integral for $x^+>0$. The modulus sign becomes superfluous in this domain of interest; therefore, it can be dispensed with.
Now, depending on whether $x^+$ is less than or greater than $\frac{P}{Q}$, the itegrand can be expanded appropriately in a series and then  integrated term by term. However, our interest in the region where $x^+$ is greater than but close to zero prompts us to work only with that expansion of the integrand which corresponds to $x^+<\frac{P}{Q}$.  This, therefore, gives us :
\begin{eqnarray}
\lambda &=&{\rm const}[(\frac{P}{Q^2}\,x^+)^\frac{1}{4}+ \sum_{n=1}^\infty 
\,{1\over n!}\,\frac{1}{4}(\frac{1}{4}-1).....(\frac{1}{4}-n+1)Q^{-\frac{1}{4}}(\frac{Q}{P})^{(n-\frac{1}{4})}
({1\over{1+4n}})(x^+)^{(n+\frac{1}{4})}]\nonumber\\
&+&{\rm const}
\label{affine parameter}
\end{eqnarray}
\noindent

This shows that the singularity occurs at a finite value of the affine parameter $\lambda$. Thus, our background spacetime, with a curvature singularity at finite affine parameter, is geodesically incomplete.

2. D-3 brane background : To continue our exercise of analysing background spacetimes, we consider as our second example the D3 brane background. 
\noindent

The geodesic equation in this case is 
\begin{equation}
\frac{d^2 x^+}{d\lambda^2} -
\frac{1}{2\,x^+}(\frac{dx^+}{d\lambda})^2 = 0. 
\end{equation}
Proceeding as in the case of our D1-D3 system we find that
\begin{equation}
y(\lambda) \equiv {dx^+\over d\lambda} = {\rm const}\:e^{-p}= {\rm const}\sqrt{x^+},
\end{equation}
\noindent
and

\begin{equation}
\lambda={\rm const}\sqrt{x^+}+{\rm const}.
\label{D3 affine parameter}
\end{equation}
This signifies geodesic incompleteness through finiteness of the affine parameter corresponding to $x^+=0$.
Moreover, we have :
\begin{equation}
R_{++}= \frac {M}{x^+}- \frac {N}{{x^+}^2}
\end{equation}
\noindent
where $M$ and $N$ are constants. 
Once again, a linear dependence of $f_{3b}$ on light-cone time yields :
\begin{eqnarray}
R_{\lambda\lambda}(\lambda) = {\rm const} (M-\frac{N}{x^+}),
\end{eqnarray}
\noindent
which blows up as $x^+ \rightarrow 0$, implying the existence of a curvature singularity at $x^+=0$.   
Thus the D3 background provides us with one more example of a cosmological background.\\
Finally, a similar analysis done with our D-string background reveals the following : 

1. The Ricci scalar $R$, and invariants such as $R^{AB}R_{AB}$, $R_{\lambda\lambda}(\lambda)$, are all regular, in fact, zero at $x^+=0$.

2. The geodesic equation gives : 
\begin{equation}
\lambda=({\rm const})\,{x^+}^{\frac{1}{4}}\,+{\rm const},
\label{}
\end{equation}
signifying a singular spacetime. The constant in the above expression can be chosen to be zero, such that $x^+=0$ corresponds to $\lambda=0$. The metric (\ref{E-frame D-string background}), expressed in terms of the affine parameter $\lambda$ and suitably scaled, reads
\begin{equation}
{ds_E}^2=2\, d{\lambda}\,dx^- - \lambda^3 \mu^2{\sum_{i=1}^{4}}{x^i}^2(d\lambda)^2 +\,\lambda\,{\sum_{a=1}^{8}}(dx^a)^2.
\label{}
\end{equation}
The non-vanishing components of the corresponding Riemann tensor diverge at $\lambda=0$, implying a curvature singularity at $\lambda=0$, where an inertial observer experiences a divergent tidal force. We are thus led to the conclusion that our D-string background too serves as yet another example of a singular spacetime.

\section{Supersymmetry Analysis}
\noindent

In this section we analyze the supersymmetry of the solutions presented above.. To do this, we write down the supersymmetry variations of the dilatino and the gravitino fields in type-II B supergravity in ten dimensions in string frame metric \cite{H} :

\begin{eqnarray}
\delta \lambda_{\pm} &=& {1\over2}(\Gamma^{\mu}\partial_{\mu}\phi \mp
\noindent{f(x^+)}^4
{1\over 12} \Gamma^{\mu \nu \rho}H_{\mu \nu \rho})\epsilon_{\pm} + {1\over
  2}e^{\phi}(\pm \Gamma^{M}F^{(1)}_{M} + {1\over 12} \Gamma^{\mu \nu
  \rho}F^{(3)}_{\mu \nu \rho})\epsilon_{\mp},
\label{dilatino}
\end{eqnarray}
\begin{eqnarray}
\delta {\Psi^{\pm}_{\mu}} &=& \Big[\partial_{\mu} + {1\over 4}(w_{\mu
  \hat a \hat b} \mp {1\over 2} H_{\mu \hat{a}
  \hat{b}})\Gamma^{\hat{a}\hat{b}}\Big]\epsilon_{\pm} \cr
& \cr
&+& {1\over 8}e^{\phi}\Big[\mp \Gamma^{\mu}F^{(1)}_{\mu} - {1\over 3!}
\Gamma^{\mu \nu \rho}F^{(3)}_{\mu \nu \rho} \mp {1\over 2.5!}
\noindent
q\Gamma^{\mu \nu \rho \alpha \beta}F^{(5)}_{\mu \nu \rho \alpha
  \beta}\Big]\Gamma_{\mu}\epsilon_{\mp},
\label{gravitino}
\end{eqnarray}
where $(\mu, \nu ,\rho)$ represent the ten dimensional space-time indices and the hat's describe the corresponding
tangent space indices. Application of the above two variation equations to the D-string solution gives us several conditions on the spinors. 

To begin with, the dilatino variation gives :
\begin{eqnarray}
\Gamma^{\hat a}\epsilon_{\pm} - \Gamma^{\hat +\hat -\hat a} 
\epsilon_{\mp} = 0.
\label{dilatino condition}
\end{eqnarray}  

In obtaining the above equation, we have imposed the condition
\begin{eqnarray}
\Gamma^{\hat +}\epsilon_{\pm} = 0.
\label{imposed condition}
\end{eqnarray} 

The gravitino variation gives the following conditions on the spinors:

\begin{eqnarray}
\delta \psi_+^{\pm} &\equiv &\partial_{+}\epsilon_{\pm}\mp{\mu\over8}
f^{-{1\over2}}_1\left((\Gamma^{\hat +\hat 1\hat 2\hat{5}\hat{6}}+\Gamma^
{\hat +\hat 3\hat 4\hat{5}\hat{6}}) +(\Gamma^{\hat +\hat 1\hat
  2\hat{7}\hat{8}}+\Gamma^{\hat +\hat 3\hat 4\hat{7}\hat{8}})\right)
\Gamma^{\hat{-}}\epsilon_{\pm}=0,\cr
& \cr
\delta \psi_-^{\pm} &\equiv& \partial_{-}\epsilon_{\pm}=0,\cr
& \cr 
\delta \psi_a^{\pm} &\equiv &\partial_{a}\epsilon_{\pm}\mp{\mu\over8}
\left((\Gamma^{\hat +\hat 1\hat 2\hat{5}\hat{6}}+\Gamma^
{\hat +\hat 3\hat 4\hat{5}\hat{6}}) +(\Gamma^{\hat +\hat 1\hat
  2\hat{7}\hat{8}}+\Gamma^{\hat +\hat 3\hat 4\hat{7}\hat{8}})\right)
\Gamma^{\hat{a}}\epsilon_{\mp}=0,\cr
&\cr
\delta \psi_p^{\pm} &\equiv &\partial_{p}\epsilon_{\pm}\mp{\mu\over8}
\left((\Gamma^{\hat +\hat 1\hat 2\hat{5}\hat{6}}+\Gamma^
{\hat +\hat 3\hat 4\hat{5}\hat{6}}) +(\Gamma^{\hat +\hat 1\hat
  2\hat{7}\hat{8}}+\Gamma^{\hat +\hat 3\hat 4\hat{7}\hat{8}})\right)
\Gamma^{\hat{p}}\epsilon_{\mp}=0,
\label{gravitino condition}
\end{eqnarray}
where we have made use of (\ref{dilatino condition}) and (\ref{imposed condition}).

Further, by using
\begin{eqnarray}
(1 + \Gamma^{\hat 1\hat 2\hat 3\hat 4})\epsilon_{\pm} = 0,
\label{third condition}
\end{eqnarray}
all the supersymmetry conditions are solved by spinors:
$\epsilon_{\pm} = \epsilon^0_{\pm}$, where
$\epsilon^0_{\pm}$ is a constant spinor.
We, therefore, conclude that our D-string solution preserves $1/8$ of the full type IIB spacetime supersymmetry. Since T-duality preserves supersymmetry, all other solutions, be they branes or bound states of branes, obtained by applying T-duality transformations, will also preserve the same amount of spacetime supersymmetry as the parent solution..

\section{Conclusion}
\noindent
In this paper we have presented the solution for a D-string in a background 
which is dependent on light-cone time and which can be viewed as a deformation 
of the pp-wave background originating from the Penrose limit of R-R 
$AdS_3\times\,S^3\times R^4$. This solution can then be used to construct other 
Dp-brane and bound state solutions in both R-R and NS-NS light-cone time 
dependent (deformed) pp-wave backgrounds by applying appropriate duality 
transformations. Our D3 brane solution and our D1-D3 bound system solution are 
just two examples of such an exercise. We notice a striking analogy between our 
solutions and those presented in a related work \cite{SSG} in that both the 
solutions, notwithstanding their very different origins, have just the constant 
part of the Green's function replaced by a linear function of time/light-cone 
time. We have, therefore, made a prescription out of this observation although 
we have not been able to supply a proof of it. An intersting discovery in this 
regard is that the backgrounds our solutions live in are all examples of cosmological backgrounds. Is our prescription, therefore, also a simple, but general, algorithm to generate readily cosmological spacetimes, at least in a classical sense? Do the singularities resolve away once quantum corrections/effects are incorporated ? The issues, in our opinion, are worth delving into. We have also studied the supersymmetric property of our D-string;  we have shown that our solution preserves one-eighth of unbroken supersymmetry. Since supersymmetry is preserved by T duality, the other solutions generated out if it will also retain the same amount of supersymmetry.

\section{Acknowledgement}
\noindent
I gratefully acknowledge my indebtedness to the late Prof. Alok Kumar for 
suggesting me the problem and providing critical inputs, and, also, for taking 
an active interest in the progress of the work. I thank S. Mukherji for a 
critical reading of the manuscript and comments that helped improve it. Thanks 
are also due to S. Banerjee for help of all sorts, too numerous to mention - not 
the least of which was help with the mathematica, and to T. Dey for help in the 
compilation of the references. I also thank S. Jain and K.L.Panigrahi for 
helpful exchanges in the early stages of the work.

\section{Appendix}
\noindent

We give below $R_{++}$ of our D1-D3 background :

\begin{eqnarray*}
R_{++}&=&[32\, \mu^2 {cos(\varphi)}^4 {X_b}^4 (\frac {{cos(\varphi)}^2\, X_b+{sin (\varphi)}^2}{{X_b^3}})^{\frac{1}{4}}\nonumber\\
&+& \,{sin(\varphi)}^4\,[X_B\,({cos (\varphi)}^2\, X_b+{sin(\varphi)}^2)]^{\frac{1}{4}}\,{{X_b}^{\prime}}^2\nonumber\\
&+&\,4 X_b\,{sin(\varphi)}^2 \,[X_B\,({cos (\varphi)}^2\, X_b+{sin(\varphi)}^2)]^{\frac{1}{4}}\,[{cos (\varphi)}^2{{X_b}^{\prime}}^2-2{sin(\varphi)}^2\,{X_b}^{\prime\prime}]\nonumber\\
&+&8\,{X_b}^3\,
[8\, \mu^2 
{cos(\varphi)}^2\,{sin(\varphi)}^2\,
(\frac {{cos(\varphi)}^2\, X_b+{sin (\varphi)}^2}{{X_b}^3})^{\frac{1}{4}}\nonumber\\
&-&{cos(\varphi)}^4 \,[X_b\,({cos (\varphi)}^2\, X_b+{sin(\varphi)}^2)]^{\frac{1}{4}}\,{{X_b}^{\prime\prime}}
]\nonumber\\
&+&\,16\,{X_b}^2\,[2\, \mu^2 \,{sin(\varphi)}^4\,(\frac {{cos(\varphi)}^2\, X_b+{sin (\varphi)}^2}{{X_b}^3})^{\frac{1}{4}}\nonumber\\
&-&\,{cos(\varphi)}^2\,{sin(\varphi)}^2\,[X_b\,({cos (\varphi)}^2\, X_b+{sin(\varphi)}^2)]^{\frac{1}{4}}\,{X_b^{\prime\prime}}]]\nonumber\\
&/&\,(\,8\,[X_b\,({cos (\varphi)}^2\, X_b+{sin(\varphi)}^2)]^{\frac{9}{4}})\nonumber\\
\end{eqnarray*}

\end{document}